\documentclass[aps,prd,showpacs,floatfix,nofootinbib,superscriptaddress,twocolumn,10pt,amsmath,array]{revtex4}
\usepackage[normalem]{ulem}
\usepackage{lineno}
\usepackage{graphicx}
\usepackage{amsmath}
\usepackage{amssymb}
\usepackage{bm}
\usepackage{color}
\usepackage{xcolor}
\usepackage[colorlinks,citecolor=blue,urlcolor=blue,linkcolor=blue]{hyperref}
\usepackage{multirow}
\usepackage{subfigure}
\usepackage{array}
\usepackage{cases}
\usepackage{booktabs}
\usepackage{balance}

\newcommand{\n}{\nonumber}

\begin{document}
\title{Synchrotron self-Compton process for constraining sub-GeV dark matter in Omega Centauri via SKA}

\author{Guan-Sen Wang}
\affiliation{Key Laboratory of Dark Matter and Space Astronomy, Purple Mountain Observatory, Chinese Academy of Sciences, Nanjing 210023, China}
\affiliation{School of Astronomy and Space Science, University of Science and Technology of China, Hefei, Anhui 230026, China}
\author{Bing-Yu Su \footnote{Corresponding author: bysu@pmo.ac.cn}}
\affiliation{Key Laboratory of Dark Matter and Space Astronomy, Purple Mountain Observatory, Chinese Academy of Sciences, Nanjing 210023, China}
\author{Yang Yu}
\affiliation{Key Laboratory of Dark Matter and Space Astronomy, Purple Mountain Observatory, Chinese Academy of Sciences, Nanjing 210023, China}
\affiliation{School of Astronomy and Space Science, University of Science and Technology of China, Hefei, Anhui 230026, China}
\author{Bo Zhang}
\affiliation{Key Laboratory of Dark Matter and Space Astronomy, Purple Mountain Observatory, Chinese Academy of Sciences, Nanjing 210023, China}
\affiliation{School of Astronomy and Space Science, University of Science and Technology of China, Hefei, Anhui 230026, China}
\author{Lei Feng \footnote{Corresponding author: fenglei@pmo.ac.cn}}
\affiliation{Key Laboratory of Dark Matter and Space Astronomy, Purple Mountain Observatory, Chinese Academy of Sciences, Nanjing 210023, China}
\affiliation{School of Astronomy and Space Science, University of Science and Technology of China, Hefei, Anhui 230026, China}

\begin{abstract}

The search for the particle identity of dark matter (DM) continues to be a primary objective in modern physics. In this field, the sub-GeV mass range of DM detection remains a crucial yet challenging window. We investigate synchrotron self-Compton (SSC) emission from electrons and positrons produced by MeV-scale DM annihilation as a novel indirect detection channel. Focusing on the globular cluster Omega Centauri and the sensitivity of the Square Kilometre Array, we derive constraints on the annihilation cross section reaching $\langle\sigma v\rangle \sim 10^{-29}\,\rm{cm}^{3}\,\rm{s}^{-1}$ in the tens-of-MeV range. Furthermore, constraints can reach below $\langle\sigma v\rangle \sim 10^{-30}\,\rm{cm}^{3}\,\rm{s}^{-1}$ under favorable parameter choices. Although the derived limits depend on the uncertain propagation parameters, the SSC channel remains competitive with existing indirect constraints over a representative range of astrophysical assumptions, establishing SSC emission as a promising probe of sub-GeV DM.

\end{abstract}


\maketitle

\section{Introduction} \label{sec:introduction}

The existence of dark matter (DM) has been firmly established by a wide range of astrophysical and cosmological observations~\cite{Cirelli:2024ssz,1939LicOB..19...41B,Jiao:2023aci,Ou:2023adg,Labini:2023fmy,Clowe:2006eq,vanWaerbeke:2000rm,Massey:2007wb,Harvey:2015hha,Robertson:2016qef,Lewis:2006fu,Hanson:2009kr,Planck:2018lbu}. However, the nature of DM still remains one of the most profound mysteries in modern physics. A well-motivated class of candidates is weakly interacting massive particles (WIMPs), whose thermal production in the early Universe naturally yields a relic abundance consistent with the observed DM density~\cite{Steigman:1984ac,Lee:1977ua,Hut:1977zn,Gunn:1978gr,Steigman:1978wqb,Davis:1981yf,Arcadi:2017kky}. In such scenarios, WIMPs can annihilate into standard model particles. This annihilation process opens a promising avenue for indirect detection, which focuses on searching for 
charged particles and photons produced by DM annihilations in astrophysical environments. 

Over the past decades, most DM indirect detection researches have focused on GeV-TeV mass range, placing strong constraints on the annihilation cross section~\cite{Abdughani:2021pdc,Cui:2016ppb,Giesen:2015ufa,Cholis:2019ejx,Cuoco:2016eej,Nguyen:2024kwy,Baring:2015sza,Fermi-LAT:2016uux,MAGIC:2016xys,Fermi-LAT:2013thd,Muru:2025vpz}. The absence of conclusive evidence in this well-explored region has motivated a strategic shift in focus toward the sub-GeV region~\cite{Knapen:2017xzo,Darme:2017glc,Boehm:2002yz,Boehm:2003hm,Boddy:2015efa,Gonzalez-Morales:2017jkx,Caputo:2022dkz,Yu:2025bwu}. However, probing this lower mass range presents distinct and significant challenges. To be specific, charged particles and photons from sub-GeV DM fall within the “MeV gap”~\cite{Carenza:2022som,Boddy:2015fsa,Knodlseder:2016pey,Engel:2022bgx}, which has long been hindered by a lack of observations. Consequently, several novel methods have recently been proposed to overcome this problem. For example, consider the reacceleration effect to enhance the energy of MeV cosmic rays and make them detectable by existing observatories~\cite{Boudaud:2016mos,Zu:2021odn,Wang:2025jhy,Su:2024hrp}. There are also researches considering secondary X-ray signals, which are produced by inverse Compton (IC) emissions from DM induced 
particles~\cite{Cirelli:2020bpc,Cirelli:2023tnx,Essig:2013goa,Balaji:2025afr}.

Beyond the mechanisms discussed above, it is also worth noting the synchrotron self-Compton (SSC) process, a specific radiative channel from MeV-scale DM annihilation. The SSC emission arises in two stages. First, relativistic $e^+ e^-$ pairs from DM annihilation generate synchrotron radiation in ambient magnetic fields. Subsequently, the SSC process occurs when these same electrons up-scatter this synchrotron photon field via inverse Compton scattering. While SSC mechanism is a standard process in high-energy astrophysics~\cite{Sari:2000zp,Ghisellini:1998it,Finke:2008pe}, the approach of using it for DM detection has not been fully explored. In fact, using SSC process to detect DM is particularly promising because, for MeV-scale DM, the peak of the SSC spectrum falls naturally within the observational window of modern radio telescopes (see Sec.~\ref{sec:SED}).

In this work, we constrain the MeV-scale DM annihilation cross section in Omega Centauri via the SSC emission. Our choice of target is the globular cluster Omega Centauri, primarily due to its status as the Milky Way's most massive cluster and a potential stripped dwarf galaxy nucleus. It is an excellent candidate for retaining a significant DM halo~\cite{Bekki:2003qw,Brown:2019whs}.
Considering the exceptional resolution and sensitivity in the radio band, Square Kilometre Array (SKA)~\cite{Braun:2019gdo} is an ideal tool to probe the subtle SSC signature from light DM annihilation. In this framework, we calculate the $e^+ e^-$ spectrum from DM annihilation in Omega Centauri, the propagation of these particles, and the resulting SSC spectrum. By comparing the predicted radio flux with the sensitivity of SKA, we present competitive upper limits on the annihilation cross section. 

This work is organized as follows: Sec.~\ref{sec:Omega_cen} briefly shows some background information of Omega Centauri, especially DM density distribution. In Sec.~\ref{sec:SED}, we explain the theoretical framework of SSC emission from DM annihilation. Our results are presented in Sec.~\ref{sec:Results}. Finally, we conclude in Sec.~\ref{sec:Conclusion}.

\section{Omega Centauri} \label{sec:Omega_cen}

As the most massive globular cluster in the Milky Way, Omega Centauri has a stellar mass of $M \sim 10^{6}\,M_{\odot}$, a half-light radius of $r_{\rm h} \sim 7\,\rm{pc}$, and is located at a distance of $d \sim 5.4\,\rm{kpc}$~\cite{Harris:1996kt,Wang:2023sxr,Wang:2021hfb,Evans:2021bsh}.
Its complex stellar populations and significant spread in metallicity suggest that it may be the remnant core of an accreted dwarf galaxy~\cite{Wang:2021hfb,Evans:2021bsh}, making it a prime candidate for retaining a substantial DM halo. Furthermore, its low gas density, typical of globular clusters, minimizes the astrophysical background radiation in the radio band. Currently, no significant diffuse radio emission from Omega Centauri has been reported~\cite{Kar:2020coz}, which makes it a clean environment for searching for a potential DM signal in the radio band.

The DM distribution in Omega Centauri is modeled using the Navarro-Frenk-White (NFW) density profile~\cite{Navarro:1996gj}
\begin{equation}
\rho(r) = \frac{\rho_{\rm s}r_{\rm s}}{r ( 1 + r/r_{\rm s})^2},
\label{eqnfw}
\end{equation}
where $\rho_{\rm s}$ is the characteristic density and $r_{\rm s}$ is the scale radius. The actual DM density distribution in Omega Centauri remains highly uncertain. In this work, we consider three representative sets of NFW parameters, summarized in Table~\ref{tab:profile}. Profile 1 is adopted as our typical parameter set for the subsequent analysis. The implications of profiles 2 and 3 will be discussed in Sec.~\ref{sec:Results} as part of a systematic uncertainty analysis.

\begin{table}[!htb]
\renewcommand\arraystretch{1.5}
\centering
\begin{tabular}{c|c|c|c}
\hline\hline
 & Profile 1 & Profile 2 & Profile 3  \\
\hline
$r_{\rm s}$~(pc) & 1.63 & 1.00 & 2.00 \\
\hline
$\rho_{\rm s}~(M_{\odot}~\rm {pc^{-3}})$ & 7650.59 & 27860.50 & 4391.65 \\
\hline\hline
\end{tabular}
\caption{The parameters of DM density distribution profiles~\cite{Wang:2021hfb}.}\label{tab:profile}
\end{table}

The synchrotron and SSC emissions further depend critically on the ambient magnetic field. The magnetic field strength in Omega Centauri is estimated to be on the order of $1\sim10\,\rm{\mu G}$~\cite{Kar:2020coz}. Here we model the magnetic field strength $B(r)$ with an exponential profile~\cite{McDaniel:2017ppt},
\begin{equation}
B(r)=B_0 e^{-r/r_{\rm c}},
\label{eqbfield}
\end{equation}
where $B_0$ is the central magnetic field strength and $r_{\rm c}$ is the core radius, which we take to be equal to the half-light radius $r_{\rm h}$~\cite{Chen:2021rea,Wang:2023sxr}. 
Furthermore, the diffusion coefficient, which is still uncertain and complicated, also affects the propagation of electrons and positrons. The associated uncertainties and our treatment of the diffusion coefficient are detailed in Sec.~\ref{sec:Results}.

\section{Synchrotron Self-Compton Emission from Dark Matter Annihilation} \label{sec:SED}
In this section, we discuss the theoretical framework of SSC process from DM annihilation, including the injection and propagation of $e^+ e^-$, and the calculation of the SSC spectrum.

We consider the annihilation of DM particles with mass $m_\chi$. In this process, the source term of $e^+ e^-$ can be described as
\begin{align}
Q(E,r)=\frac{\langle\sigma v\rangle\rho_{\chi}^2(r)}{2m_{\chi}^2}\frac{{\rm{d}}N}{{\rm{d}}E},
\label{esource}
\end{align}
where $\langle\sigma v\rangle$ is the cross section of DM annihilation, the DM density $\rho_{\chi}(r)$ is given by the NFW profile as described in Eq.\,\eqref{eqnfw}, and ${\rm{d}}N/{\rm{d}}E$ is the $e^{+}e^{-}$ spectrum produced by one annihilation event. Here we take the spectrum from PPPC4~\cite{Cirelli:2010xx} and extrapolate it to the sub-GeV mass range of our study.

Assuming these $e^{+}e^{-}$ propagate through diffusion and undergo continuous energy loss, their equilibrium density is described by
\begin{align}
\frac{\partial}{\partial t}\frac{\partial n_e}{\partial E}=&\nabla\left[D(E,r){\nabla} \frac{\partial n_e}{\partial E}\right]\nonumber\\
&+\frac{\partial}{\partial E}\left[b(E,r)\frac{\partial n_e}{\partial E}\right]+Q(E,r),
\label{edensity}
\end{align}
where $D(E,r)$ and $b(E,r)$ model the diffusion coefficient and energy loss term, respectively. In this work, we solve it numerically using the Green's function method~\cite{Colafrancesco:2005ji,Ginzburg:1964} with the RX-DMFIT package~\cite{McDaniel:2017ppt}. 

Since the diffusion coefficient is still not precisely known, we use a simplified form, expressing it as a power law in energy~\cite{Colafrancesco:2005ji,DAMPE:2022jgy,Wang:2023sxr,McDaniel:2017ppt}, 
\begin{align}
D(E)=D_0(E/E_0)^{1/3},
\label{D0}
\end{align}
where $E_0 = 1\,{\rm GeV}$ and $D_0$ represents the diffusion constant. For Omega Centauri, $D_0$ is estimate to be $10^{26}$--$10^{30}~\rm{cm^2~s^{-1}}$~\cite{Kar:2020coz,Wang:2023sxr}.

The energy loss term $b(E,r)$, detailed as
\begin{align}
b(E,r)&=b_{\rm IC}(E)+b_{\rm Synch}(E,r)+b_{\rm Coul}(E)+b_{\rm Brem}(E) \nonumber \\
&=(b_{\rm ICSL}^0+b_{\rm ICCMB}^0)E^2+b_{\rm Synch}^0B(r)^2E^2 \nonumber\\
  &\quad+b_{\rm Coul}^0n_e\left[1+{\lg}\left(\frac{E/m_e}{n_e}\right)/75\right] \nonumber\\
  &\quad+b_{\rm Brem}^0n_e\left[{\lg}\left(\frac{E/m_e}{n_e}\right)+0.36\right],
\label{energyloss}
\end{align}
includes contributions from synchrotron loss $b_{\rm Synch}(E,r)$, IC loss $b_{\rm IC}(E)$, Coulomb loss $b_{\rm Coul}(E)$, and bremsstrahlung loss $b_{\rm Brem}$~\cite{McDaniel:2017ppt}. Here, the IC energy loss includes contributions from both the cosmic microwave  background (CMB) and the stellar radiation field. The thermal electron density $n_e$ is taken as $0.05 \, \rm{{cm}^{-3}}$~\cite{Freire:2001wq}, typical for globular clusters, and the energy loss coefficients (all in units of $10^{-16}\, \rm{GeV/s}$) are~\cite{Colafrancesco:2005ji,Profumo:2010ya}
\begin{align}
b_{\rm syn}^0\approx0.0254,&\quad b_{\rm ICCMB}^0 \approx 0.25,\n\\
b_{\rm ICSL}^0 \approx 6.08,\quad b_{\rm brem}^0&\approx 1.51,\quad b_{\rm Coul}^0 \approx 6.13.
\label{bcoefficient}
\end{align}
In fact, the SSC process itself introduces an additional energy loss term for the electrons. However, our calculation shows that this contribution is negligible within the parameter space considered in this work. Consequently, it has not been included in our work. A detailed discussion is provided later in this section.

With the electron distribution obtained from \eqref{edensity}, the synchrotron radiation spectrum can be calculated by
\begin{align}
j_{\rm syn}(\nu,r)=2\int_{m_e}^{m_{\chi}}{\rm{d}}E\,\frac{{\rm{d}}n_e}{{\rm{d}}E}(E,r)P_{\rm syn}(\nu,E,r),
\label{jsynch}
\end{align}
where $P_{\rm syn}(\nu,E,r)$, the standard synchrotron power, describes the synchrotron photon production rate at frequency $\nu$ from an electron with energy $E$. It also changes along with the position $r$ where the synchrotron emission happens. More details of $P_{\rm syn}(\nu,E,r)$ can be found in~\cite{McDaniel:2017ppt}. Assuming the synchrotron photons propagate freely in the Universe, we can calculate the equilibrium photon density based on
\begin{align}
n_{\rm syn}(\nu,r)=\frac{1}{4\pi r^2 c}\int_{0}^{r}4\pi R^2 j_{\rm syn}(\nu,R)\,{\rm{d}}R.
\label{nesynch}
\end{align}
Here we neglect the influence caused by the second step of SSC process, which is not expected to be significant. 

At this point, using the electron distribution from \eqref{edensity} and the photon distribution from \eqref{nesynch}, we are ready to calculate SSC emission spectrum through
\begin{align}
j_{\rm SSC}(\nu,r)=2\int_{m_e}^{m_{\chi}}{\rm{d}}E\,\frac{{\rm{d}}n_e}{{\rm{d}}E}(E,r)P_{\rm SSC}(\nu,E,r).
\label{jSSC}
\end{align}
Here, $P_{\rm SSC}(\nu,E,r)$, though termed the SSC power, is computed using the standard inverse Compton scattering formula. It gives the photon production rate at frequency $\nu$ for a single electron of energy $E$. More details can be found in ~\cite{McDaniel:2017ppt}.

\begin{figure}[htb]
\centering
\includegraphics[width=\linewidth]{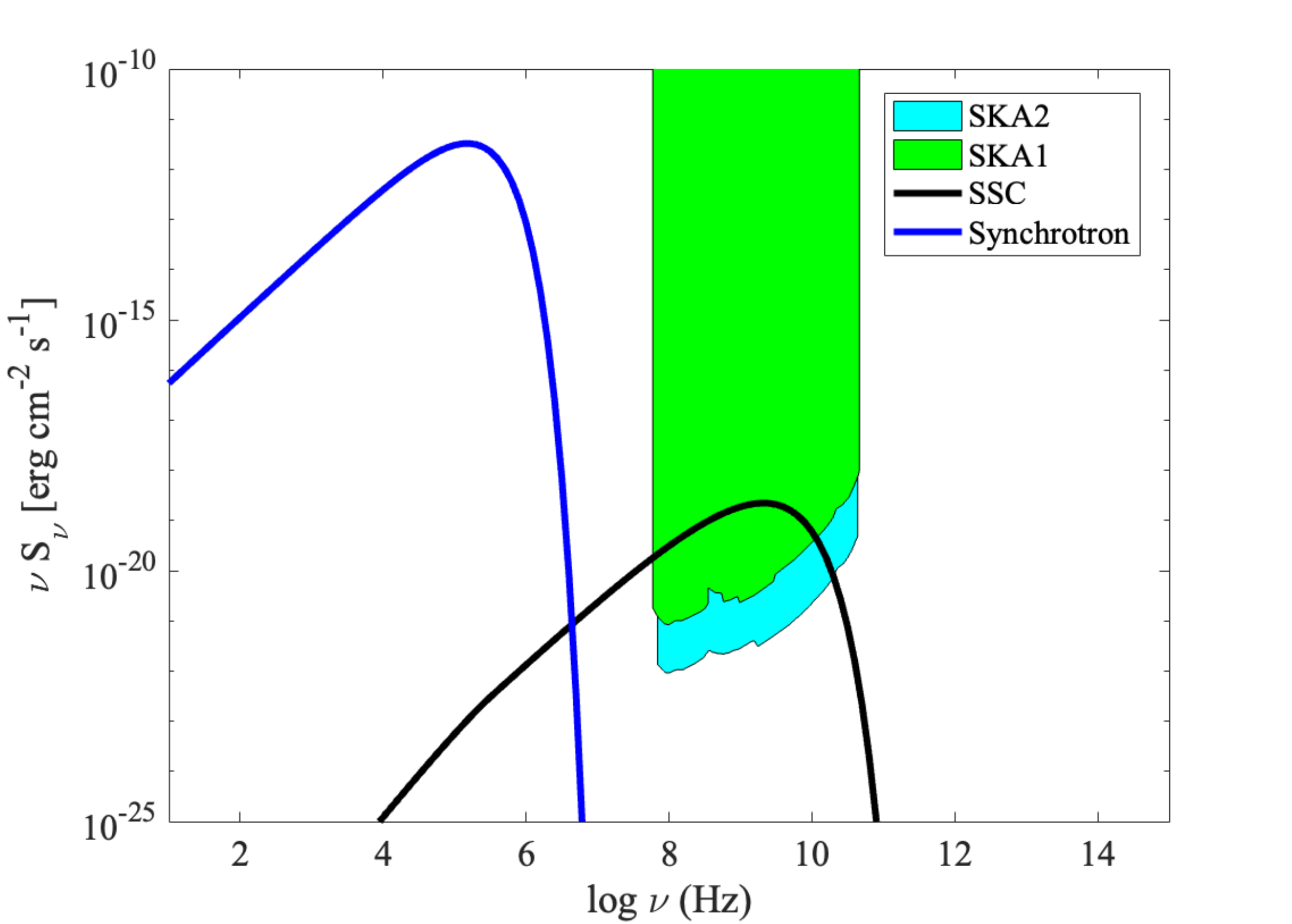}
\caption{The energy flux of SSC emission (black line) compared with the energy flux of synchrotron (blue line). Here we choose $m_\chi = 50~\rm{MeV}$, $D_0 = 10^{28}~\rm {cm^{2}~s^{-1}}$, $\langle\sigma v\rangle=10^{-28} ~\rm{cm^{3}~s^{-1}}$, $B_0=5~\rm {\mu G}$. The green and cyan regions represent the sensitivity of SKA phase 1 and phase 2 with $100~{\rm h}$ of observation, respectively~\cite{Braun:2019gdo,Wang:2023sxr}.}
\label{fig:flux}
\end{figure}

Finally, we can compute the flux density spectrum through
\begin{align}
\nu S(\nu)=\nu \int_{\Omega}{\rm{d}}\Omega \int_{\rm los} j_{\rm SSC}(\nu,r)\,{\rm{d}}l,
\label{flux}
\end{align}
In Fig.~\ref{fig:flux}, we present the energy flux of SSC emission as the black line, and that of synchrotron as the blue line for comparison. The shaded region indicates the sensitivity of SKA with $100~{\rm h}$ observation~\cite{Braun:2019gdo,Wang:2023sxr}. Since SKA is still under construction, we adopt this nominal sensitivity as a benchmark for assessing the detectability of the predicted SSC signal.
Here we choose DM density profile 1 from Table~\ref{tab:profile}, with the DM particle mass $m_\chi = 50~\rm{MeV}$, the diffusion coefficient $D_0 = 10^{28}~\rm {cm^{2}~s^{-1}}$, the DM annihilation $\langle\sigma v\rangle=10^{-28} ~\rm{cm^{3}~s^{-1}}$, and the magnetic field $B_0=5~\rm {\mu G}$.
Our calculation indicates that for the 50 MeV DM mass, the peak of the SSC spectrum lies within the observational frequency range of SKA, and the predicted SSC flux exceeds the sensitivity threshold. Therefore, the SSC mechanism offers a unique method for DM indirect detection, particularly within the challenging sub-GeV region. Moreover, it is observed that the synchrotron energy flux significantly exceeds that of the SSC emission component. This supports our earlier simplification of neglecting the energy loss from the SSC process, as its contribution is subdominant in the overall energy budget.

\section{Results} \label{sec:Results}
\begin{figure}[htb]
\centering
\includegraphics[width=\linewidth]{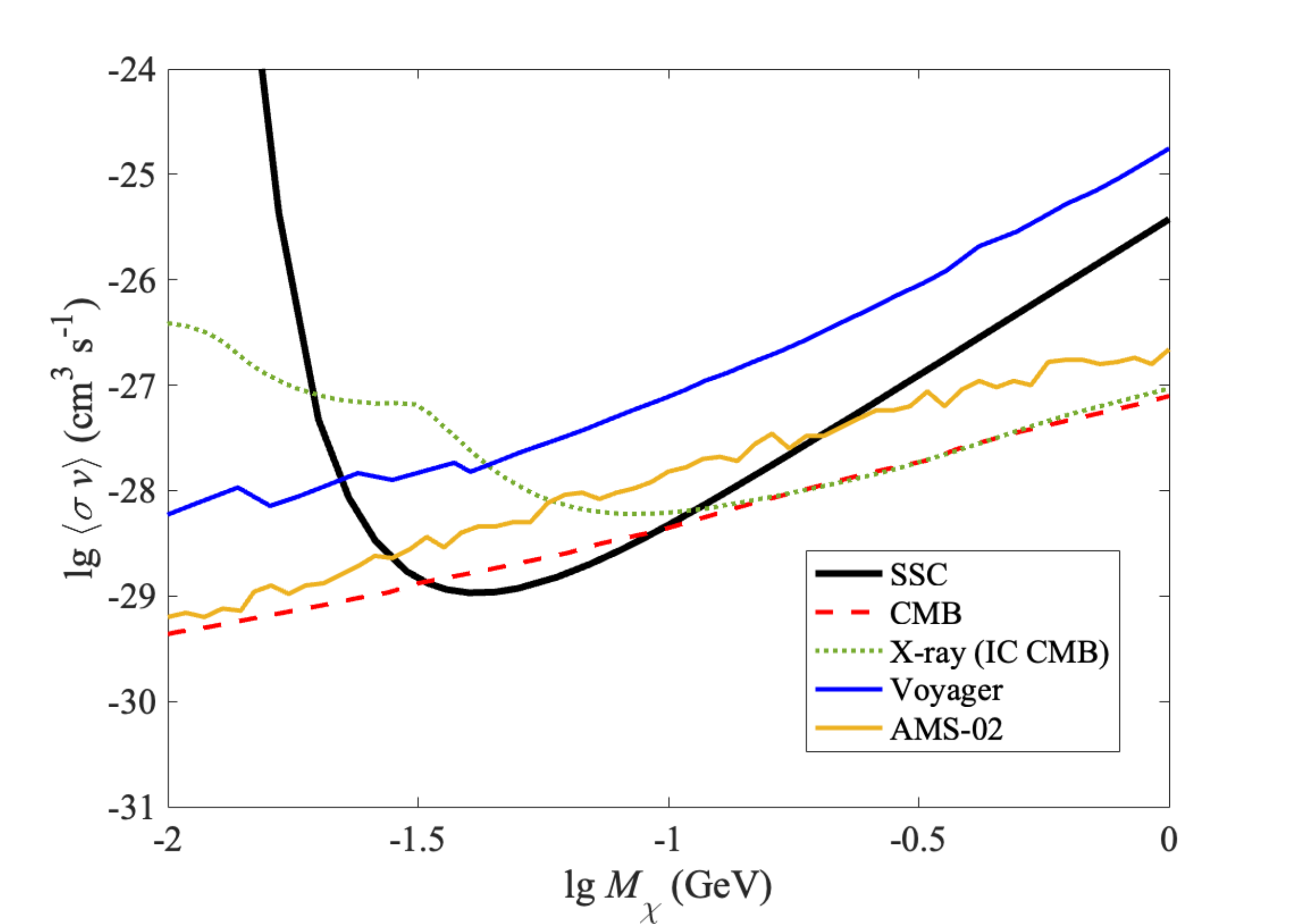}
\caption{Our constraints of DM annihilation cross section with $D_0 = 10^{28}~\rm {cm^{2}~s^{-1}}$ and $B_0=5~\rm {\mu G}$ (black solid line), compared with limits from CMB data~\cite{Slatyer:2015jla,Leane:2018kjk} (red dashed line); Voyager cosmic-ray data~\cite{Boudaud:2016mos} (blue solid line); AMS-02 cosmic-ray data~\cite{Wang:2025jhy} (yellow solid line) and secondary x-ray emissions~\cite{Cirelli:2020bpc,Cirelli:2023tnx} (green dotted line).}
\label{fig:sigmav}
\end{figure}

In this section, we compare the energy flux of SSC emission from DM with the sensitivity of SKA to constrain the DM annihilation cross section. We compute the minimum annihilation cross section for which the predicted radio flux equals to the sensitivity of SKA. We adopt profile 1 from Table~\ref{tab:profile} as the fiducial DM density distribution, along with a diffusion coefficient of $D_0 = 10^{28}\,\rm{cm}^{2}\,\rm{s}^{-1}$ and a magnetic field strength of $B_0 = 5\,\mu\rm{G}$. The results are shown in Fig.~\ref{fig:sigmav} as the black solid line. We adopt the sensitivity at $1~\rm{GHz}$, which is $2.6\times10^{-7}~\rm{Jy}$ for simplicity~\cite{Braun:2019gdo,Wang:2023sxr}. Also, we present results from other established methods for comparison. These include the red dashed line from CMB data~\cite{Slatyer:2015jla,Leane:2018kjk}; limits from cosmic-ray observations with Voyager~\cite{Boudaud:2016mos} (blue solid line) and AMS-02~\cite{Wang:2025jhy} (yellow solid line); and bounds derived from the search for secondary x-ray emissions~\cite{Cirelli:2020bpc,Cirelli:2023tnx} (green dotted line). A comparative analysis reveals that our constraints derived from the SSC channel are particularly strong within the $30$–$100\,\rm{MeV}$ mass range, and exceed $\langle\sigma v\rangle\sim10^{-29} ~\rm{cm^{3}~s^{-1}}$ in the several tens of MeV range.

\begin{figure}
\centering 
\subfigure[]
{\includegraphics[width=1\linewidth]{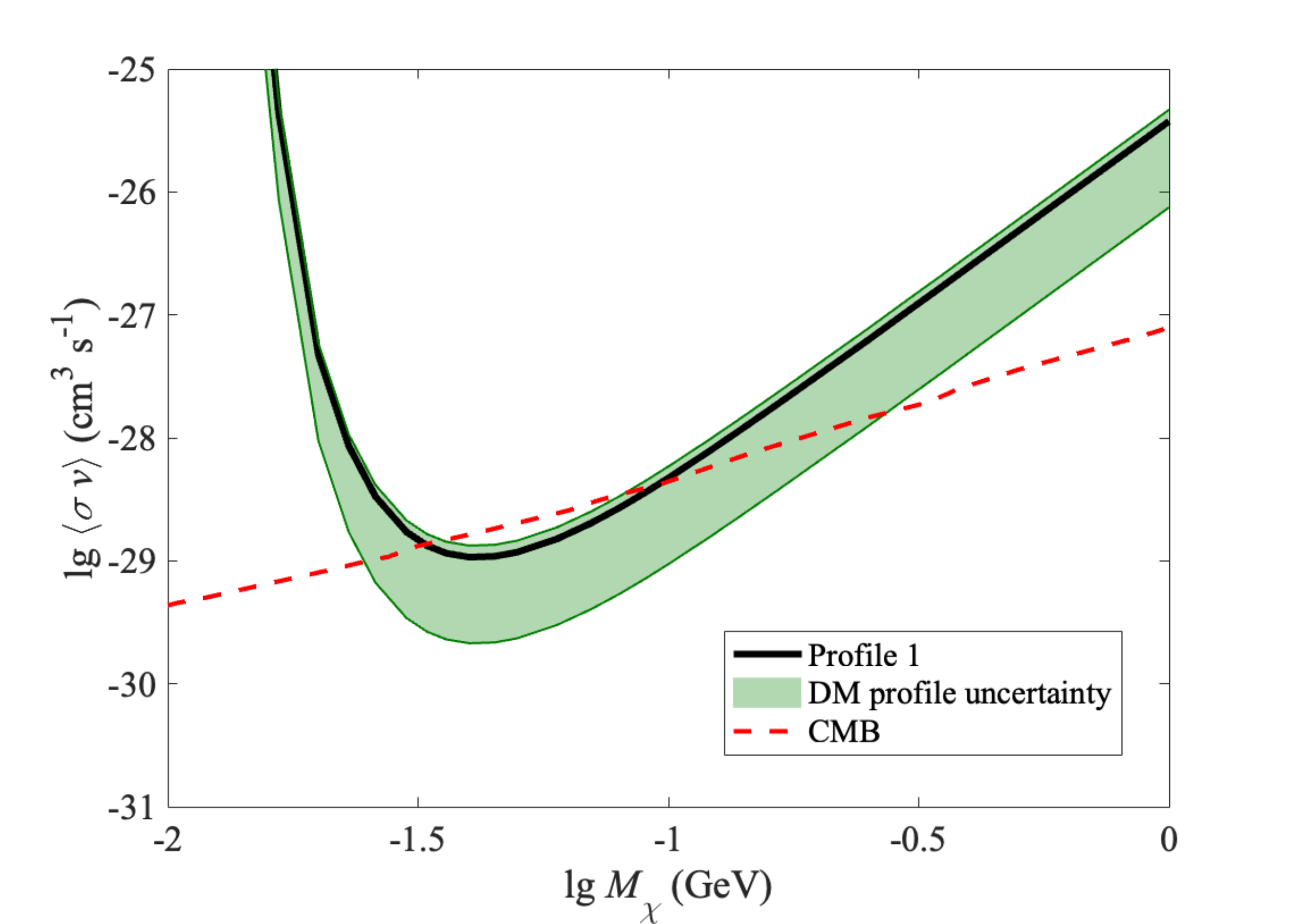}\label{fig:DMdensity}}\\
\subfigure[]{\includegraphics[width=1\linewidth]{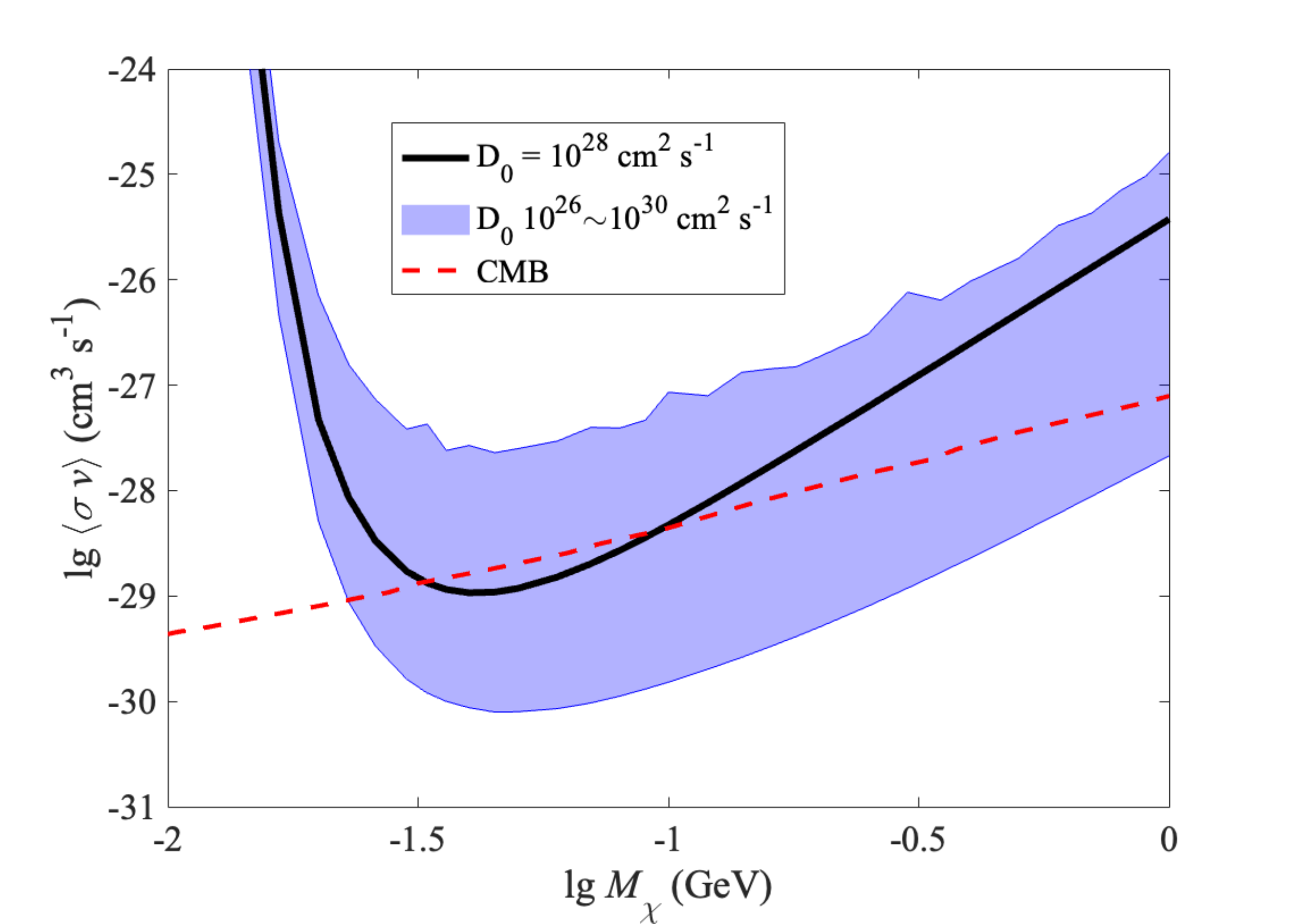}\label{fig:D0}} \\
\subfigure[]
{\includegraphics[width=1\linewidth]{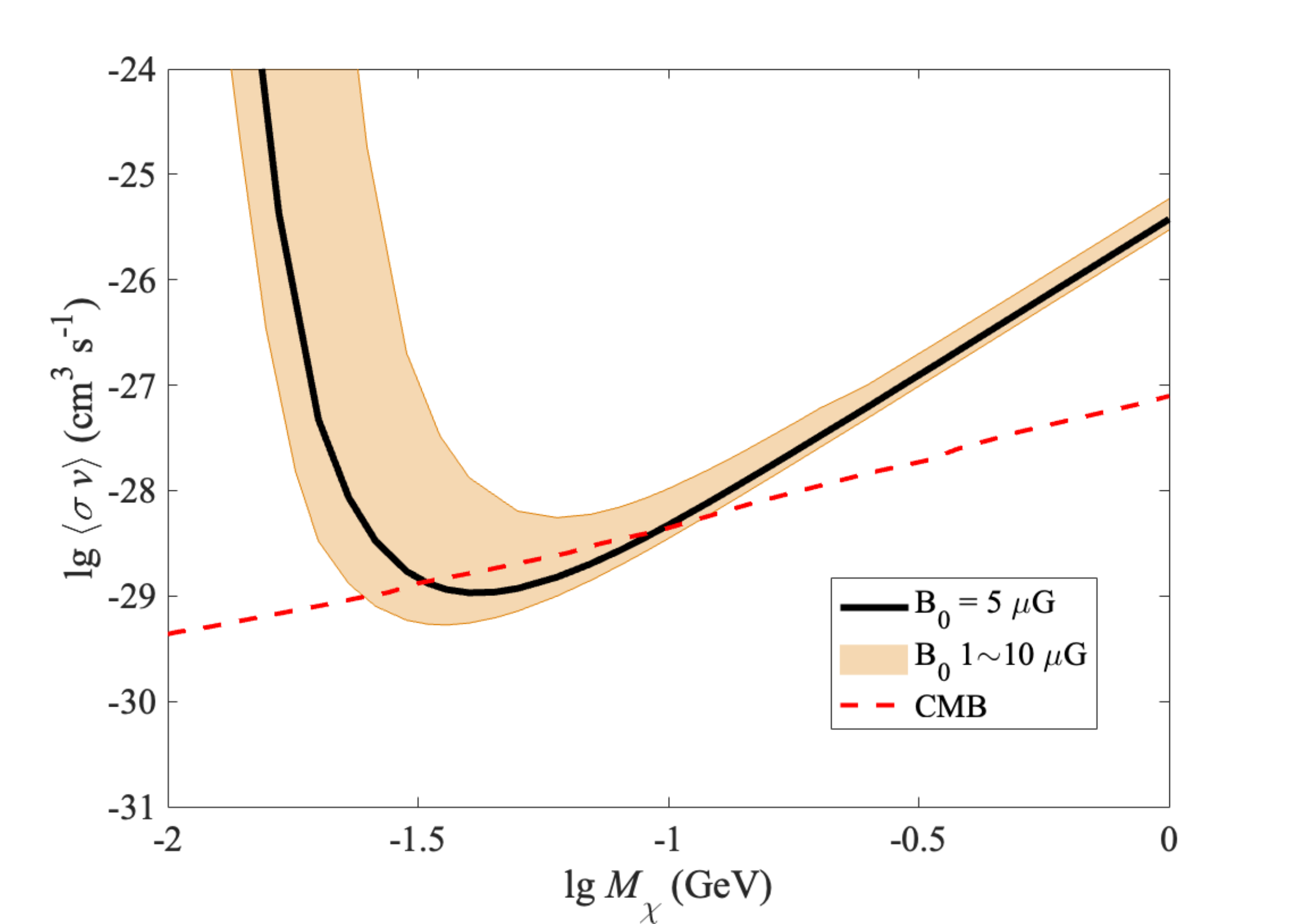}\label{fig:B0}}
\caption{Our fiducial constraints of DM annihilation cross section (black solid line), together with the effects of astrophysical uncertainties. The (a) green, (b) blue, and (c) yellow bands show variations due to the DM density profile (profiles 2, 3), diffusion coefficient ($D_0=10^{26}$--$10^{30}\,\rm{cm}^{2}\,\rm{s}^{-1}$), and magnetic field ($B_0=1$--$10\,\mu\rm{G}$), respectively.}
\label{fig:uncertainty}
\end{figure}

Moreover, our results are subject to uncertainties in several key astrophysical parameters: the DM density profile $\rho(r)$ in Omega Centauri, the magnetic field strength $B$, and the diffusion coefficient $D_0$. To quantify their impact, we present a dedicated uncertainty analysis in Fig.~\ref{fig:uncertainty}. Here the solid black line represents our fiducial results shown in Fig.~\ref{fig:sigmav} before.
The variations induced by each parameter are illustrated as shaded bands:
The green band shows the effect of varying the DM density profile, with its lower and upper edges corresponding to Profile 2 and Profile 3, respectively. The blue band illustrates the uncertainty from the diffusion coefficient, with the lower and upper edges corresponding to $D_0 = 10^{26}$ and $10^{30}\,\rm{cm}^{2}\,\rm{s}^{-1}$, respectively. The yellow band presents the range due to the magnetic field, varying from $B_0 =1\,\mu\rm{G}$ (upper edge)to $B_0 = 10\,\mu\rm{G}$ (lower edge).

In general, although these uncertainties can significantly affect the quantitative limits, the SSC method remains a competitive probe of sub-GeV DM over a representative range of astrophysical parameters. Moreover, the SSC constraints are less sensitive to the uncertainty in the magnetic field $B_0$ than to that in the diffusion coefficient $D_0$. This is because the magnetic field influences only the first stage of the SSC process (synchrotron emission), making the final constraints more robust against variations in $B_0$.

\section{Conclusion} \label{sec:Conclusion}

In this study, we investigated the feasibility of using SSC radiation as a novel indirect detection channel to search sub-GeV DM particles. We find that within the 30–100 MeV range, SSC method delivers stringent constraints down to $\langle\sigma v\rangle \sim 10^{-29}\,\rm{cm}^{3}\,\rm{s}^{-1}$, outperforming existing bounds. Furthermore, our analysis of uncertainties from key astrophysical parameters reveals that the derived constraints are least sensitive to the DM density profile, followed by the magnetic field $B$, and most sensitive to the diffusion coefficient. The overall uncertainty bands also indicate that significantly tighter constraints could be achieved under more favorable astrophysical conditions, with the potential to surpass $\langle\sigma v\rangle \sim 10^{-30}\,\rm{cm}^{3}\,\rm{s}^{-1}$ in optimistic scenarios. Overall, the SSC process offers a unique and highly promising observational window for probing the MeV gap that is challenging for traditional methods.

\acknowledgments

This work is supported by the National Key R\&D Program of China (Grant No. 2022YFF0503304), the National Natural Science Foundation of China (Grants No. 12373002,  No. 12220101003, and No. 11773075), and the Youth Innovation Promotion Association of Chinese Academy of Sciences (Grant No. 2016288).

\bibliography{refprd}
\bibliographystyle{aapmrev4-2}

\end{document}